\documentclass[12pt]{article}
\usepackage{amsmath,amssymb}

\usepackage{graphicx}
\usepackage{wrapfig}

\usepackage{hyperref}
\usepackage{cite}

\def\beq{\begin{eqnarray}}
\def\eeq{\end{eqnarray}}

\def\o{\over}

\newcommand{\be}{\begin{equation}}
\newcommand{\ee}{\end{equation}}
\newcommand{\bea}{\begin{eqnarray}}
\newcommand{\eea}{\end{eqnarray}}
\newcommand{\bg}{\begin{gather}}
\newcommand{\eg}{\end{gather}}
\newcommand{\bseq}{\begin{subequations}}
\newcommand{\eseq}{\end{subequations}}

\renewcommand{\ln}{\mathop{\rm ln}\nolimits}

\def\be{\begin{eqnarray}}
\def\ee{\end{eqnarray}}
\def\lb{\label}

\begin{document}

\title{\textbf{Entanglement  entropy  of round spheres}}

\vspace{1cm}
\author{ \textbf{
 Sergey N. Solodukhin$^\sharp$ }} 

\date{}
\maketitle

\begin{center}
  \hspace{-0mm}
  \emph{ Laboratoire de Math\'ematiques et Physique Th\'eorique }\\
  \emph{Universit\'e Fran\c cois-Rabelais Tours F\'ed\'eration Denis Poisson - CNRS, }\\
  \emph{Parc de Grandmont, 37200 Tours, France} \\
\end{center}

{\vspace{-11cm}
\begin{flushright}
\end{flushright}
\vspace{11cm}
}



\begin{abstract}
\noindent { We propose that the logarithmic term in the entanglement entropy computed in  a conformal field theory  for a $(d-2)$-dimensional round sphere in Minkowski  spacetime is
identical to the logarithmic term in the entanglement entropy of extreme black hole. The near horizon geometry of the latter is $H_2\times S_{d-2}$. For a scalar field this proposal is checked by direct    calculation.  We comment on relation of this and earlier calculations to the ``brick wall'' model of 't Hooft. The case of generic 4d conformal field theory is discussed.}
\end{abstract}

\vskip 2 cm
\noindent
\rule{7.7 cm}{.5 pt}\\
\noindent 
\noindent
\noindent ~~~$^{\sharp}$ {\footnotesize e-mail: Sergey.Solodukhin@lmpt.univ-tours.fr}


\newpage

\section{ Introduction}
\setcounter{equation}0
Entanglement entropy  \cite{BS}, \cite{Srednicki:1993im}, \cite{Frolov:1993ym}  is defined with respect to a surface $\Sigma$ by tracing over the modes that reside inside the surface. The correlations that exist in the system make the entropy to be determined by  geometry of the surface and, to leading order in dimension $d>2$, by the area of $\Sigma$. For a recent review on the entanglement entropy in free quantum field theory see \cite{Casini:2009sr}.  
The concept of  entanglement entropy is very natural  when applied to black hole horizons and is thought to lay in the origin of the black hole entropy, although the details of this relation are yet to be unveiled \cite{Dvali:2008jb}. 

A somewhat related approach was introduced by 't Hooft \cite{'tHooft:1984re} and is known as the ``brick wall'' model. In this approach 
one considers the entropy of thermal excitations of quantum field modes that propagate just outside the black hole horizon. The density of these modes
becomes infinite when one approaches the horizon. In order to regularize this divergence 't Hooft introduced a brick wall, an imaginary boundary that stays at distance $\epsilon$ from the actual horizon. The entropy calculated in this approach is proportional to the area of the horizon and diverge
when $\epsilon$ is taken to zero. This behavior of the entropy  is similar to that of the entanglement entropy. However, we note that  this approach is applicable only when there is a (black hole or Rindler) horizon. 

Recently, the interest in entanglement entropy has been revived due to proposal of Ryu and Takayanagi \cite{Ryu:2006bv} that entanglement entropy in a conformal field theory which has a dual geometric description can be calculated in a completely geometric fashion as area of a minimal surface in $(d+1)$-dimensional anti-de Sitter spacetime. This proposal has been checked in many particular cases and a complete agreement with the known results has been found.

A particular focus of the current research has been made on the study of the logarithmic terms in the entanglement entropy. In dimensions higher than 2
these terms were first found in the entropy of black holes \cite{Solodukhin:1994yz}, \cite{Solodukhin:1994st}. In particular, the entanglement
entropy  of Reissner-Nordstrom black hole of mass $M$ and electric charge $Q$ for a scalar field in 4 spacetime dimensions takes the form \cite{Solodukhin:1994st}
\be
S_{RN}={A\o 48\pi \epsilon^2}-({1\o 18}-{M\o 15 r_+})\ln\epsilon\; ,
\lb{1}
\ee
where $A=4\pi r_+^2$ is the area of horizon, $r_+=M+\sqrt{M^2-Q^2}$, and $\epsilon$ is an UV cutoff.
For extreme black hole $M=Q$ and we have that
\be
S_{ext}={A\o 48\pi \epsilon^2}+{1\o 90}\ln\epsilon\; ,
\lb{2}
\ee
as it was found in \cite{Solodukhin:1994st}. In fact, this formula should be understood as the entropy in the universal extremal limit of non-extremal geometries as demonstrated in \cite{Mann:1997hm}.

From a different  perspective the logarithmic terms in the entropy of extreme black holes in four dimensions has been recently discussed in \cite{Banerjee:2010qc} where one suggests that the correspondence to the microscopic theory requires vanishing of these terms in $N=4$ supegravity, the cancellation even occurs for the matter multiplet taken separately.

In another recent development the logarithmic term in the entropy was calculated for a round sphere in flat Minkowski space-time \cite{Solodukhin:2008dh}, \cite{Casini:2010kt}, \cite{Dowker:2010nq}.
For a generic 4d conformal field theory characterized by the conformal anomalies of type $A$ and $B$ the entropy was shown to be  \cite{Solodukhin:2008dh}
\be
S={A(\Sigma)\o 48\pi\epsilon^2}+A\pi^2\ln\epsilon\; ,
\lb{3}
\ee
so that the logarithmic term is determined only by the anomaly of type A.
 In a theory with $n_s$ particles of spin $s$ one finds \cite{Duff:1993wm} (the contributions of fields of spin $3/2$ and $2$ can be obtained from table 2 on p.180 of the book of Birrell and Davies \cite{BD})
\be
&&A={1\o 90\pi^2}(n_0+11n_{1/2}+62n_1+0n_{3/2}+0n_2)\; , \nonumber \\
&&B={1\o
30\pi^2} (n_0 + 6n_{1/2} + 12n_1-{233\o 6}n_{3/2}+{424\o 3} n_2)\; .
\lb{4}
\ee
The derivation of formula (\ref{3}) in \cite{Solodukhin:2008dh} was based on conformal invariance of the logarithmic term and the correspondence to the holographic description of entanglement entropy proposed in \cite{Ryu:2006bv}. In the case of $N=4$ superconformal field theory holographically dual to 
supergravity on $AdS_5$ one has $A=B$ and formula (\ref{3}) is the entropy predicted in the prescription of \cite{Ryu:2006bv}.

For a scalar field one has $A={1\o 90\pi^2}$ and we find that the logarithmic term in (\ref{3}) is identical to the logarithmic term in the entropy of extreme black hole (\ref{2}). In this note we show that this is not a coincidence. In fact, we show that the logarithmic terms are identical in two apparently rather different
situations: $(d-2)$-sphere in flat Minkowski $d$-dimensional spacetime and spherical horizon of a static $d$-dimensional extreme black hole.
In the latter case the near horizon geometry is the product $H_2\times S_{d-2}$, where $H_2$ is hyperbolic space, what in the physics literature is
known as Euclidean $AdS_2$. In particular, in four spacetime dimensions, we claim that formula (\ref{3}) gives the entanglement entropy of extreme 4d black hole in a generic conformal field theory thus extending the result of  \cite{Solodukhin:2008dh}.

In a recent paper \cite{Casini:2010kt} Casini and Huerta have extended the result (\ref{3})  for a scalar field to the entropy of $(d-2)$-dimensional sphere.
They have managed to calculate the logarithmic term in even dimensions up to $d=14$. In a subsequent paper \cite{Dowker:2010nq} Dowker 
has given an alternative calculation and obtained the logarithmic term in dimensions $d=15$ and $d=16$. Clearly, the algorithmic procedures proposed in 
\cite{Casini:2010kt} or \cite{Dowker:2010nq} allow one to determine the logarithmic term for any value of $d$. The result of this note suggests that the same values should be attributed to the entanglement entropy of $d$-dimensional extreme black hole.

It should be noted that the issue of the logarithmic term in entanglement  entropy has become   principally important.
This is due to claims made in \cite{Schwimmer:2008yh} that the holographic description \cite{Ryu:2006bv} of entanglement entropy is not general enough
and cannot give the right description of the entropy if surface $\Sigma$ is characterized by non-trivial extrinsic curvature.
In particular, it was claimed that the discrepancy manifests for a round sphere in flat spacetime. Now, as the direct  calculations of the entropy for a sphere in flat spacetime are available analytically \cite{Casini:2010kt}, \cite{Dowker:2010nq} and numerically \cite{Lohmayer:2009sq} and these calculations confirm the formula (\ref{3}) one has no  reason  to doubt that the holographic proposal gives the right predictions for entanglement entropy.  On the contrary, there is accumulating evidence that the entanglement entropy is yet another example of the power of the holographic AdS/CFT correspondence.

\section{ General structure of UV divergences}
\setcounter{equation}0
Before proceeding to the particular calculations we would like to specify the general structure of entanglement entropy.
In $d$ spacetime dimensions entanglement entropy is presented in a form of the Laurent series with respect to 
UV cutoff $\epsilon$ (for $d=4$ see  	\cite{Solodukhin:2008dh})
\be
S={s_{d-2}\o \epsilon^{d-2}}+{s_{d-4}\o \epsilon^{d-4}}+.. +{s_{d-2n}\o \epsilon^{d-2n}}+..+s_0\ln\epsilon+s(g)\; ,
\lb{2.1}
\ee
where $s_{d-2}$ is proportional to the area of surface $\Sigma$. All other terms in the expansion 
(\ref{2.1}) can be presented as integrals over surface $\Sigma$ of local quantities constructed in terms of Riemann curvature of the spacetime and the extrinsic curvature of surface $\Sigma$. Since nothing should depend on the direction of  vectors normal to $\Sigma$, the integrands in expansion (\ref{2.1}) should be even powers of extrinsic curvature. Thus, since the integrands are even in derivatives then only terms $\epsilon^{d-2n}$, $n=0, 1, 2, ..$ may appear in expansion (\ref{2.1}). If dimension $d$ is even then there also may appear a logarithmic term $s_0$. This is consistent with the fact that $s_0$ is the surface term in the integrated conformal anomaly, the latter is non-vanishing only if dimension $d$ is even.  

\section{ Conformal transformation}
\setcounter{equation}0
Consider now a sphere of radius $R$ in flat Minkowski spacetime. We can choose a spherical coordinate system $(\tau,r,\theta^i)$ so that the surface $\Sigma$ 
is defined as $\tau=0$ and $r=R$, and variables $\theta^i\; , \; i=1,..,d-2$ are the angular coordinates on $\Sigma$. The $d$-metric reads
\be
ds^2=d\tau^2+dr^2+r^2\gamma_{ij}(\theta)d\theta^id\theta^j\; ,
\lb{3.1}
\ee
where $\gamma_{ij}(\theta)$ is metric on $(d-2)$ sphere of unite radius.  Metric (\ref{3.1}) is conformal to the metric
\be
ds_{ext}^2={R^2\o r^2}(d\tau^2+dr^2)+R^2\gamma_{ij}(\theta)d\theta^id\theta^j\; ,
\lb{3.2}
\ee
which describes the product of two-dimensional hyperbolic space $H_2$ with coordinates $(\tau,r)$ and the sphere $S_{d-2}$.
Note the both spaces, $H_2$ and $S_{d-2}$, have the same radius $R$. Metric (\ref{3.2}) describes the spacetime which appears in the extremal
limit of $d$-dimensional static black hole. In the hyperbolic space $H_2$ we can choose a polar coordinate system $(\rho,\phi)$  with the center at point $r=R$,
\be
r={R\o \cosh\rho-\sinh\rho\cos\phi}~,~~\tau={R\sinh\rho\sin\phi\o \cosh\rho-\sinh\rho\cos\phi}\; ,
\lb{3.3}
\ee
(for small $\rho$ one has that $r=R+\rho\cos\phi~,~~\tau=\rho\sin\phi$ as in the polar system in flat spacetime)
so that the metric takes the form
\be
ds_{ext}^2=R^2(d\rho^2+\sinh^2\rho d\phi^2)+R^2\gamma_{ij}(\theta)d\theta^id\theta^j\; .
\lb{3.4}
\ee
In this coordinate system the black hole horizon is at $\rho=0$. This is the point where surface $\Sigma$ was located in the old coordinates, $r=R,\ \tau=0$. The  metric (\ref{3.4}) is obtained from a static metric
\be
&&ds^2_{BH}=g(x)dt^2+g^{-1}(x)dx^2+x^2\gamma_{ij}(\theta)d\theta^id\theta^j\; ,\nonumber \\
&&g(x)=g_0(x-R)+R^{-2}(x-R)^2+O((x-R)^2)\; ,
\lb{3.5}
\ee
which describes black hole with horizon at $x=R$, by taking the extremal limit $g_0\rightarrow 0$ in a manner originally proposed by Zaslavsky \cite{Zaslavsky:1997ha} and applied to the entropy calculation in \cite{Mann:1997hm}. 
In Zaslavsky's method one reaches the extremal limit whilst remaining in the topological class of the corresponding non-extremal geometries.
The limiting geometry (\ref{3.4}) is characterized by finite temperature determined by  the $2\pi$ periodicity in angular coordinate $\phi$.

\section{ Entropy of extreme black hole and the logarithmic terms}
\setcounter{equation}0
Calculation of entropy in metric (\ref{3.4}) goes along standard lines by first introducing a conical singularity at the horizon (this is achieved by making coordinate $\phi$ $2\pi\alpha$-periodic), evaluating the effective action of the field operator $\cal D$ in question on manifold with conical singularity and then differentiating the result with respect to $\alpha$. One uses the heat kernel in order to evaluate the effective action. For $d=4$ this procedure for metric (\ref{3.4}) was  done in \cite{Mann:1997hm}. The result is given by equation (25) in \cite{Mann:1997hm}.  Here we should add only few minor modifications. First of all the field operator we consider is the conformal scalar field operator. In $d$ dimensions it takes the form
\be
{\cal D}=-(\nabla^2+E)~,~~E=-{(d-2)\o 4(d-1)}R_{(d)}\; ,
	\lb{4.1}
	\ee
	where $R_{(d)}$ is Ricci scalar. For metric (\ref{3.4}) one has that $R_{(d)}=R^{-2}(-2+(d-2)(d-3))=R^{-2}(d-1)(d-4)$ and hence (we set radius $R=1$)
	\be
	E=-{(d-2)(d-4)\o 4}\; .
	\lb{4.2}
	\ee
	 In $d=4$ the non-minimal coupling in (\ref{4.1}) vanishes, this is the case considered in \cite{Mann:1997hm}.
The other modification is due the fact that metric (\ref{3.4}) is direct product of $H_2$ and $S_{d-2}$ while in \cite{Mann:1997hm} the spherical part was $S_2$.  With all these modifications taken into account the entropy  for metric (\ref{3.4}) is given by
\be
S_{ext}={1\o 4\sqrt{\pi}}\int^\infty_{\epsilon^2/R^2}{ds\o s^{3/2}}k_H(s)\Theta_{d-2}(s)e^{-s/4}e^{sE}\; ,
\lb{4.3}
\ee
where function $k_H(s)$ is given by expression (see equation (26) in \cite{Mann:1997hm})
\be
k_H(s)=\int_0^\infty dy {\cosh y\o \sinh^2 y}(1-{2y\o \sinh(2y)})e^{-y^2/s}\; .
\lb{4.4}
\ee
Function $\Theta_{d-2}(s)$ is the trace of heat kernel of Laplace operator $-\nabla^2$ on $(d-2)$-dimensional sphere of unite radius.
This function is presented in the form of expansion
\be
\Theta_{d-2}(s)={\Omega_{d-2}\o (4\pi s)^{(d-2)/2}}\left(1+(d-2)(d-3)\sum_{n=1}^\infty a_{2n}s^n\right)\; ,  
\lb{4.5.1}
\ee
where $\Omega_{d-2}={2\pi^{(d-1)/2}\o \Gamma((d-1)/2)}$ is the area of a unit radius sphere $S_{d-2}$.
The first few coefficients in this expansion can be calculated using the results reported in \cite{HK},
\be
&&a_2={1\o 6}~,~~a_4={(5d^2-27d+40)\o 360}~,~~
a_6={(35d^4-392d^3+1699d^2-3322d+2520)\o 45360}\; ,\lb{4.5.2} \\
&&a_8={(6125d^6-106575d^5+781865d^4-3100197d^3+7106558 d^2-9051960 d+5124672)\o 7938000}\; . \nonumber
\ee
 On the other hand, the function $k_H(s)$ (\ref{4.4}) is represented by the series
\be
k_H(s)=\sqrt{\pi s}({1\o 3}-{1\o 20}s+{17\o 1120}s^2-{29\o 4480}s^3+{1181\o 337920}s^4-{1393481\o 615014400}s^5+{763967\o 447283200}s^6+..)\; 
\lb{4.6}
\ee
Combining everything together and using (\ref{4.3}) we find that  in dimension $d=4$  the logarithmic term is $s_0={1\o 90}$ as found in \cite{Mann:1997hm}, in dimension
$d=6$ one has $s_0=-{1\o 756}$ and in dimension $d=8$ one finds $s_0={23\o 113400}$ in agreement with \cite{Casini:2010kt}. In odd dimensions the logarithmic term is zero as 
predicted by the arguments presented in section 2. The calculation of $s_0$ for higher values of $d$ requires the knowledge of more terms in the expansion
(\ref{4.5.1}).

\section{ Relation to  ``brick wall'' model of 't Hooft }
\setcounter{equation}0
In metric (\ref{3.4}) the coordinate $\phi$ plays the role of the Euclidean time. By making one more conformal transformation one may transform
(\ref{3.4}) to the form in which $g_{\phi\phi}=1$. This is the so-called optical metric, it takes the form
\be
&&ds^2_{opt}=d\phi^2+\sinh^{-2}\rho(d\rho^2+\gamma_{ij}(\theta)d\theta^id\theta^j )\nonumber \\
&&=d\phi^2+dy^2+\sinh^2y\; \gamma_{ij}(\theta)d\theta^id\theta^j\; ,
\lb{5.1}
\ee
where in the second line we introduced a new  variable $y$, $\sinh \rho=1/\sinh y$, which changes from $0$ to $\infty$. The $(y,\theta^i)$ part of metric (\ref{5.1}) describes 
$(d-1)$-dimensional anti-de Sitter space $H_{d-1}$. The horizon $\Sigma$ which stayed at $\rho=0$ now lies at infinity of the anti-de Sitter, $y=\infty$. The total spacetime described by (\ref{5.1}) is the direct product $S_1\times H_{d-1}$.
The Ricci scalar of the metric (\ref{5.1}) is that of anti-de Sitter, $R_{(d)}=-(d-1)(d-2)$
We note in passing that it is a general feature of black hole horizons that in the optical metric there always appears asymptotically anti-de Sitter space, the fact that allows to formulate a duality similar to the AdS/CFT correspondence which acts at the horizon \cite{Sachs:2001qb}. 

The optical metric is natural to use when one describes the wave  propagation on the black hole background. In our case of conformal scalar field the relevant wave operator (in the Euclidean signature) is
\be
{\cal D}=-\partial_\phi^2+ \tilde{\cal D}_{d-1}~,~~\tilde{\cal D}_{d-1}=-(\tilde{\nabla}^2+\tilde{E})~,~~\tilde{E}={(d-2)^2\o 4}\; ,
\lb{5.2}
\ee
where $\tilde{\cal D}_{d-1}$ is operator acting on the anti-de Sitter space $H_{d-1}$. 

In the ``brick wall'' model of 't Hooft \cite{'tHooft:1984re} (see also \cite{brickwall1}) one considers a thermal gas of modes described by (Minkowski signature version of) operator (\ref{5.2})
and evaluates the corresponding entropy. Alternatively (see for example \cite{brickwall2}), in the Euclidean part integral approach one evaluates the free energy at temperature $T=1/2\pi$ of the field system
with wave operator (\ref{5.2}) on the background of optical metric. The entropy obtained in this approach is (see \cite{brickwall2}) 
\be
S_{BW}=T^{d-1}V_{d-1} \;( d\pi^{-d/2}\Gamma({d\o 2})\zeta (d)+..)\; ,
\lb{5.3}
\ee
where $..$ stands for terms due to curvature of $H_{d-1}$,
is divergent due to infinite volume $V_{d-1}$ of
the anti-de Sitter space $H_{d-2}$.  These divergences can be regularized by putting a boundary at some large value of $y=-\ln\epsilon$ that corresponds  to a boundary (the brick wall) at some small distance $\rho=\epsilon$ from the horizon in the coordinate system $(\rho, \phi, \theta^i)$.
In fact the divergences with respect to $\epsilon$ are precisely the UV divergences 
that can be seen by introducing the Pauli-Villars regularization as was demonstrated in \cite{Demers:1995dq}. The brick wall divergences then are replaced by the UV divergences with respect to the Pauli-Villars regulator. The ``brick wall'' entropy  is thus characterized by UV divergences which have structure similar to that of entanglement entropy. 

In the present context we note that the ``brick wall'' entropy is exactly the entropy which was calculated by Casini and Huerta in \cite{Casini:2009sr}.
The wave operator (\ref{5.2}) is the operator which appears in their calculation (they normalize the temperature to $1$ so that $\tilde E$ in their case is multiplied by $4\pi^2$). The terms $..$ in (\ref{5.3}) are explicitly calculated in \cite{Casini:2009sr}. The logarithmic term in the entropy comes from
the small $\epsilon$ expansion of the volume  $V_{d-1}$ and coincides as we have observed in this paper with the logarithmic term in the entanglement entropy
of extreme black hole (\ref{3.4}). It is curious that by conformal transformations the ``brick wall'' entropy (\ref{5.3}) and the entropy of extreme black hole (\ref{4.3}) are related to the entropy of a round sphere in flat spacetime.

\section{ Logarithmic term in generic 4d conformal field theory}
\setcounter{equation}0
As we propose in this paper, for a generic 4d conformal field theory characterized by anomalies of type A and B the logarithmic term in the entropy of 4d extreme
black hole (the near horizon geometry is $H_2\times S_2$) is the same as in the entropy of round sphere in flat spacetime,
\be
s_0^{ext}(A,B)=A\pi^2\; .
\lb{6.1}
\ee
In  fact, this formula  is a consequence of general expression for logarithmic term on an arbitrary gravitational background  obtained in \cite{Solodukhin:2008dh}, see equation (2.11) in \cite{Solodukhin:2008dh}. 
This term thus depends only on the anomaly of type $A$. This is contrary to the case of the Schwarszchild black hole (which  is another example of black hole geometry characterized by just one dimensionfull parameter), when both A and B anomalies contribute to the logarithmic term in the entropy 
\be
s_0^{sch}(A,B)=(A-B)\pi^2\; .
\lb{6.2}
\ee
For a scalar field this formula can be checked directly using (\ref{1}) and (\ref{4}). Again, for a generic conformal field theory it is a consequence of more general expression (2.11) in \cite{Solodukhin:2008dh}. The logarithmic term (\ref{6.2}) is related to the UV finite term computed in \cite{Fursaev:1994te}\footnote{We believe there is an overall  sign error in \cite{Fursaev:1994te} in the logarithmic term.}.
The combination $(A-B)$  vanishes in $N=8$ and $N=4$ supergravities and in $N=4$ super-Yang-Mills
theory (see  \cite{Duff:1993wm}). In the latter case this is a consequence of identity $A=B$.
On the other hand, the anomaly of type $A$, taken separately,  is not vanishing in these theories that should result in appearance of a non-vanishing logarithmic term in the entropy of extreme black hole in those theories\footnote{In $N=4$ superconformal gauge field theory the logarithmic term was computed holographically in \cite{Solodukhin:2006xv}  for an arbitrary static black hole, for the particular cases of extreme black hole and the Schwarzschild black hole this result agrees with (\ref{6.1}) and (\ref{6.2}).}. 
Apparently, the situation is  more subtle  if the background fluxes are present\footnote{I thank A. Sen for comments on this point.}. As is shown in  \cite{Banerjee:2010qc} the effect of the background flux on the fermionic fields results in changing the sign of their contribution to the anomaly and the logarithmic term in the entropy. The logarithmic term then may vanish for a matter multiplet consisting of scalars, vectors and fermions as demonstrated in \cite{Banerjee:2010qc}. On the other hand, as can be seen from (\ref{4}) the fields of spin $s=2$ (graviton) and $s=3/2$ (gravitino) do not contribute to the anomaly of type A. Thus the corresponding logarithmic term in the entropy vanishes for these fields. This observation agrees with the anticipation of paper \cite{Banerjee:2010qc} that the logarithmic term should also vanish in the gravitational sector of $N=4$ supegravity
(so that the total logarithmic term vanishes as required by the correspondence to the microscopic results, for references and further motivations see \cite{Banerjee:2010qc}).
The effect of the fluxes in the gravitational sector of $N=4$ supergravity  however remains to be understood.

\section{ Conclusions}
\setcounter{equation}0
In this note a round $(d-2)$-sphere appears in three different situations: in flat spacetime, as boundary  of anti-de Sitter in the Euclidean space $S_1\times  H_{d-1}$ and as horizon of extreme black hole with geometry $H_2\times S_{d-2}$.  All three spacetimes are related by conformal transformations.
The logarithmic term in entanglement entropy of a round sphere is conformal invariant and thus can be calculated in three different approaches.
Respectively this was done in \cite{Solodukhin:2008dh},  \cite{Casini:2010kt} and in the present note which generalizes an earlier calculation of 
\cite{Mann:1997hm}. In  work of Dowker \cite{Dowker:2010nq} one maps, by means of a conformal transformation, $d$-dimensional flat spacetime to $d$-sphere. This is yet another way to formulate the problem.
The results obtained in these different approaches  agree. The agreement  shows that predictions based on the holographic description \cite{Ryu:2006bv} of entanglement entropy are correct.
Entanglement entropy in flat spacetime has a better chance to be measured in an experiment. This indirectly would give information on the entropy of black holes
and may provide us with an experimental evidence for  the conjectured AdS/CFT correspondence.

\bigskip

The idea of this paper has crystallized while the author was staying at the Sirenis Seaview Country Club (Ibiza). The constant support and encouragement from
ACEP is gratefully acknowledged.
The useful discussions with Dmitry Nesterov are appreciated. I thank Ashoke Sen for his interesting comments.


\begin{thebibliography}{999}

{\frenchspacing \parskip=2mm

\bibitem{BS}
  L.~Bombelli, R.K.~Koul, J.H.~Lee and R.D.~Sorkin,
  ``A Quantum Source Of Entropy For Black Holes,''
  Phys.\ Rev.\ D {\bf 34}, 373 (1986).

\bibitem{Srednicki:1993im}
  M.~Srednicki,
  ``Entropy and area,''
  Phys.\ Rev.\ Lett.\  {\bf 71} (1993) 666
  [arXiv:hep-th/9303048].
  
  
\bibitem{Frolov:1993ym}
  V.~P.~Frolov and I.~Novikov,
  ``Dynamical origin of the entropy of a black hole,''
  Phys.\ Rev.\  D {\bf 48}, 4545 (1993)
  [arXiv:gr-qc/9309001].

  
\bibitem{Casini:2009sr}
 H.~Casini and M.~Huerta,
  ``Entanglement entropy in free quantum field theory,''
  arXiv:0905.2562 [hep-th].
 
\bibitem{Dvali:2008jb}
  G.~Dvali and S.~N.~Solodukhin,
  ``Black Hole Entropy and Gravity Cutoff,''
  arXiv:0806.3976 [hep-th]; and work in progress

 
\bibitem{'tHooft:1984re}
  G.~'t Hooft,
  ``On The Quantum Structure Of A Black Hole,''
  Nucl.\ Phys.\  B {\bf 256}, 727 (1985).

\bibitem{Ryu:2006bv}
  S.~Ryu and T.~Takayanagi,
  ``Holographic derivation of entanglement entropy from AdS/CFT,''
  Phys.\ Rev.\ Lett.\  {\bf 96}, 181602 (2006)
  [arXiv:hep-th/0603001];
  
S.~Ryu and T.~Takayanagi,
  ``Aspects of holographic entanglement entropy,''
  JHEP {\bf 0608} (2006) 045
  [arXiv:hep-th/0605073].

\bibitem{Solodukhin:1994yz}
  S.~N.~Solodukhin,
  ``The conical singularity and quantum corrections to entropy of black hole,''
  Phys.\ Rev.\  D {\bf 51}, 609 (1995)
  [arXiv:hep-th/9407001].
  
\bibitem{Solodukhin:1994st}
  S.~N.~Solodukhin,
  ``On 'Nongeometric' contribution to the entropy of black hole due to quantum
  corrections,''
  Phys.\ Rev.\  D {\bf 51}, 618 (1995)
  [arXiv:hep-th/9408068].

\bibitem{Mann:1997hm}
  R.~B.~Mann and S.~N.~Solodukhin,
  ``Universality of quantum entropy for extreme black holes,''
  Nucl.\ Phys.\  B {\bf 523}, 293 (1998)
  [arXiv:hep-th/9709064].

\bibitem{Banerjee:2010qc}
  S.~Banerjee, R.~K.~Gupta and A.~Sen,
  ``Logarithmic Corrections to Extremal Black Hole Entropy from Quantum Entropy
  Function,''
  arXiv:1005.3044 [hep-th].
  
\bibitem{Solodukhin:2008dh}
  S.~N.~Solodukhin,
  ``Entanglement entropy, conformal invariance and extrinsic geometry,''
  Phys.\ Lett.\  B {\bf 665}, 305 (2008)
  [arXiv:0802.3117 [hep-th]].

\bibitem{Casini:2010kt}
  H.~Casini and M.~Huerta,
  ``Entanglement entropy for the n-sphere,''
  arXiv:1007.1813 [hep-th].

\bibitem{Dowker:2010nq}
  J.~S.~Dowker,
  ``Hyperspherical entanglement entropy,''
  arXiv:1007.3865 [hep-th].



\bibitem{Duff:1993wm}
  M.~J.~Duff,
  ``Twenty years of the Weyl anomaly,''
  Class.\ Quant.\ Grav.\  {\bf 11}, 1387 (1994)
  [arXiv:hep-th/9308075].
  
\bibitem{BD} N.~D.~Birrell and P.~C.~W.~Davies, {\it Quantum Fields in Curved Space} (Cambridge Univ. Press, New York 1982).

\bibitem{Schwimmer:2008yh}
  A.~Schwimmer and S.~Theisen,
  ``Entanglement Entropy, Trace Anomalies and Holography,''
  Nucl.\ Phys.\  B {\bf 801}, 1 (2008)
  [arXiv:0802.1017 [hep-th]].

\bibitem{Lohmayer:2009sq}
  R.~Lohmayer, H.~Neuberger, A.~Schwimmer and S.~Theisen,
  ``Numerical determination of entanglement entropy for a sphere,''
  Phys.\ Lett.\  B {\bf 685}, 222 (2010)
  [arXiv:0911.4283 [hep-lat]].
  
\bibitem{Zaslavsky:1997ha}
  O.~B.~Zaslavsky,
  ``Geometry of nonextreme black holes near the extreme state,''
  Phys.\ Rev.\  D {\bf 56}, 2188 (1997)
  [Erratum-ibid.\  D {\bf 59}, 069901 (1999)]
  [arXiv:gr-qc/9707015].


\bibitem{HK}
  D.~V.~Vassilevich,
  ``Heat kernel expansion: User's manual,''
  Phys.\ Rept.\  {\bf 388}, 279 (2003)
  [arXiv:hep-th/0306138];
  
  I.~G.~Avramidi,
  ``The Covariant technique for the calculation of the heat kernel asymptotic
  Phys.\ Lett.\  B {\bf 238}, 92 (1990).

  
\bibitem{Sachs:2001qb}
  I.~Sachs and S.~N.~Solodukhin,
  ``Horizon holography,''
  Phys.\ Rev.\  D {\bf 64}, 124023 (2001)
  [arXiv:hep-th/0107173].


\bibitem{brickwall1}
 L.~Susskind and J.~Uglum,
  ``Black hole entropy in canonical quantum gravity and superstring theory,''
  Phys.\ Rev.\  D {\bf 50}, 2700 (1994)
  [arXiv:hep-th/9401070];

D.~N.~Kabat and M.~J.~Strassler,
  ``A Comment on entropy and area,''
  Phys.\ Lett.\  B {\bf 329}, 46 (1994)
  [arXiv:hep-th/9401125].


\bibitem{brickwall2}
J.~L.~F.~Barbon,
  ``Horizon divergences of fields and strings in black hole backgrounds,''
  Phys.\ Rev.\  D {\bf 50}, 2712 (1994)
  [arXiv:hep-th/9402004];
  
S.~P.~de Alwis and N.~Ohta,
  ``Thermodynamics Of Quantum Fields In Black Hole Backgrounds,''
  Phys.\ Rev.\  D {\bf 52}, 3529 (1995)
  [arXiv:hep-th/9504033].


\bibitem{Demers:1995dq}
  J.~G.~Demers, R.~Lafrance and R.~C.~Myers,
  ``Black hole entropy without brick walls,''
  Phys.\ Rev.\  D {\bf 52}, 2245 (1995)
  [arXiv:gr-qc/9503003].


\bibitem{Fursaev:1994te}
  D.~V.~Fursaev,
  ``Temperature And Entropy Of A Quantum Black Hole And Conformal Anomaly,''
  Phys.\ Rev.\  D {\bf 51}, 5352 (1995)
  [arXiv:hep-th/9412161].


\bibitem{Solodukhin:2006xv}
  S.~N.~Solodukhin,
  ``Entanglement entropy of black holes and AdS/CFT correspondence,''
  Phys.\ Rev.\ Lett.\  {\bf 97}, 201601 (2006)
  [arXiv:hep-th/0606205].


} 

\end{thebibliography}
\end{document}